\documentclass[preprint,journal]{vgtc}            

\onlineid{1307}

\ieeedoi{10.1109/TVCG.2024.3456315}

\vgtccategory{Research}

\vgtcpapertype{application/design study}

\usepackage[table,dvipsnames]{xcolor}
\usepackage{graphicx}
\usepackage{tikz}
\usepackage{accsupp}

\newcommand{\IconBIMView}[0]{\includegraphics[height=0.25cm]{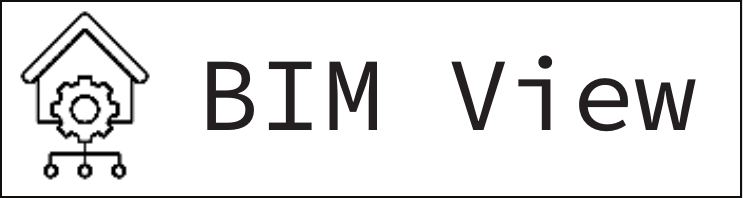}\hspace{5pt}}
\newcommand{\IconSpaceView}[0]{\includegraphics[height=0.25cm]{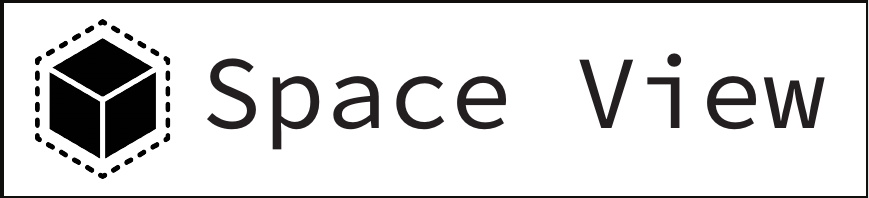}\hspace{5pt}}
\newcommand{\IconElementView}[0]{\includegraphics[height=0.25cm]{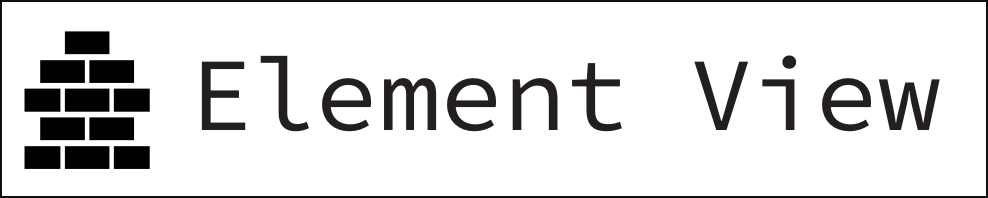}\hspace{5pt}}
\newcommand{\IconRelationshipView}[0]{\includegraphics[height=0.25cm]{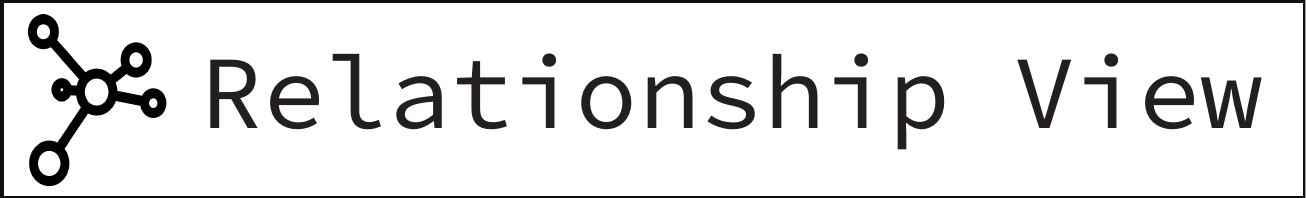}\hspace{5pt}}
\newcommand{\IconBEMView}[0]{\includegraphics[height=0.25cm]{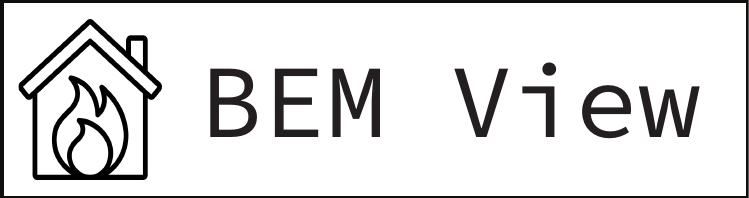}\hspace{5pt}}

\definecolor{EvaluationGood}{RGB}{1, 133, 113}
\definecolor{EvaluationMiddle}{RGB}{128, 205, 193}
\definecolor{EvaluationLow}{RGB}{223, 194, 125}
\definecolor{EvaluationBad}{RGB}{244, 165, 130}

\definecolor{mybrown}{RGB}{223,136,0}
\definecolor{myblue}{RGB}{31,120,180}
\definecolor{mygreen}{RGB}{51,160,44}

\title{BEMTrace: Visualization-driven approach for deriving \\Building Energy Models from BIM}

\author{%
    \authororcid{Andreas Walch}{0000-0002-4567-7942},
    \authororcid{Attila Szabo}{0000-0002-1322-8704}
    \authororcid{Harald Steinlechner}{0000-0002-5179-7799}
    \authororcid{Thomas Ortner}{0000-0002-9373-6409},
  \\
    \authororcid{Eduard Gröller}{0000-0002-8569-4149}, and
    \authororcid{Johanna Schmidt}{0000-0002-9638-6344}
}

\authorfooter{
  \item
  	Andreas Walch, Attila Szabo, Harald Steinlechner, and Johanna Schmidt are with VRVis Zentrum für Virtual Reality und Visualisierung Forschungs-GmbH, Vienna, Austria.\\
  	E-mail: [walch|szabo|steinlechner|schmidt]@vrvis.at.
  \item 
        Thomas Ortner is an independent researcher, Vienna, Austria\\
        E-mail: thomas@ortner.fyi.
  \item
  	Eduard Gröller is with TU Wien, Vienna, Austria.\\
  	E-mail: groeller@cg.tuwien.ac.at.
}

\abstract{Building Information Modeling (BIM) describes a central data pool covering the entire life cycle of a construction project. Similarly, Building Energy Modeling (BEM) describes the process of using a 3D representation of a building as a basis for thermal simulations to assess the building’s energy performance. This paper explores the intersection of BIM and BEM, focusing on the challenges and methodologies in converting BIM data into BEM representations for energy performance analysis. BEMTrace integrates 3D data wrangling techniques with visualization methodologies to enhance the accuracy and traceability of the BIM-to-BEM conversion process. Through parsing, error detection, and algorithmic correction of BIM data, our methods generate valid BEM models suitable for energy simulation. Visualization techniques provide transparent insights into the conversion process, aiding error identification, validation, and user comprehension. We introduce context-adaptive selections to facilitate user interaction and to show that the BEMTrace workflow helps users understand complex 3D data wrangling processes.
}

\keywords{BIM, BEM, BIM-to-BEM, 3D Data Wrangling, 3D selections, Visualization for trust building}

\teaser{
    \begin{tikzpicture}[outer sep=0, inner sep=0, font=\sffamily\small]
        \BeginAccSupp{method=pdfstringdef, ActualText={We show the five different views of our application. All views are 3D views. The first one on the left, the BIM View, renders the BIM model data, which is the input data. Here, a 3D rendering of a building is shown. The Space View to its right shows only the building's spaces (or rooms), rendered as colored blocks. The following view, the Element View, renders building elements like walls. The Relationship View, right next to it, is the most complex part of our application. It shows the connection areas between rooms, defined as space boundaries, which are rendered as colored planes. As a last view in our pipeline, the BEM View renders the final BEM model.}}
            \node[anchor=north west] at (0,0) {
                \includegraphics[width=\textwidth]{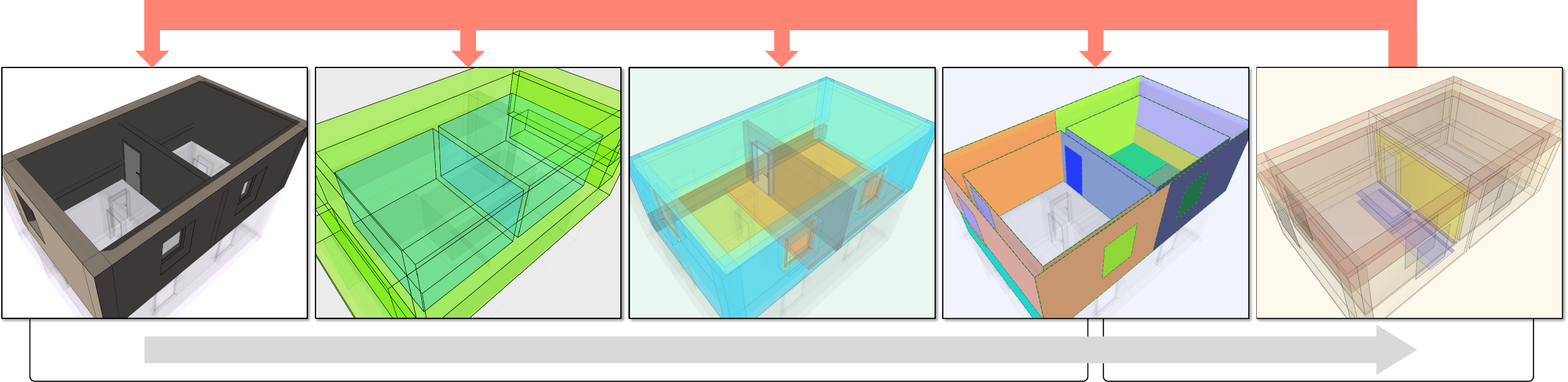}};
            \node[text=white] at (8,-0.18) {Context-adaptive selections};
            \node at (8.5,-3.655) {BIM-to-BEM conversion};
            \node at (6,-4.21) {BIM World};
            \node at (13.7,-4.21) {BEM World};
            \node[anchor=north west] at (.015,-2.97) {\includegraphics[height=0.35cm]{pics/icon-bimview}};
            \node[anchor=north west] at (3.3,-2.97) {\includegraphics[height=0.35cm]{pics/icon-spaceview}};
            \node[anchor=north west] at (6.59,-2.975) {\includegraphics[height=0.35cm]{pics/icon-elementview}};
            \node[anchor=north west] at (9.875,-2.975) {\includegraphics[height=0.35cm]{pics/icon-relationshipview}};
            \node[anchor=north west] at (13.17,-2.97) {\includegraphics[height=0.35cm]{pics/icon-bemview}};
        \EndAccSupp{}
    \end{tikzpicture}
    \caption{With BEMTrace, we visually support the data curation process from a building information model (BIM) to a building energy model (BEM), denoted as \emph{BIM-to-BEM conversion}. Users are provided with different views of the data to help them understand the complex data conversion process. The views are either associated with BIM data (\emph{BIM World}), with BEM data (\emph{BEM World}), or with the transition between these two worlds (\emph{Relationship View}). \emph{Context-adaptive selections} support users in navigating through the different views and exploring different aspects of the data in more detail.}
  \label{fig:teaser}
}

\graphicspath{{figs/}{figures/}{pictures/}{images/}{./}} 

\usepackage{tabu}                      
\usepackage{booktabs}                  
\usepackage{lipsum}                    
\usepackage{mwe}                       

\usepackage{mathptmx}                  

\begin{document}


\firstsection{Introduction}

\maketitle

The construction industry underwent a digital transformation with a fundamental shift towards data-driven and collaborative tools and processes in construction projects~\cite{Chen23}. One of the most significant shifts was developing and establishing the concept of \emph{Building Information Modeling (BIM)}. While architects and engineers have already widely adopted computer-aided design (CAD), BIM in construction is a step further. BIM involves the creation of a 3D digital model, where each object can hold additional information about its components, materials, and their relationships~\cite{BorrKKB15}. BIM offers numerous advantages in the construction process, including improved collaboration, error reduction, and the possibility of applying enhanced visualization techniques that allow stakeholders to visualize the project more effectively.

Planning a new building involves various design choices and decisions. Heating ventilation and air conditioning (HVAC) systems provide buildings with fresh air and adjust the heating and cooling load. Evolving from successfully applying BIM in the building design process, engineers and designers started to establish computational procedures that involve simulating and analyzing a building's energy performance based on a BIM representation. In essence, this \emph{Building Energy Modeling (BEM)} process describes a virtual simulation of a building's energy consumption, thermal behavior, and overall efficiency, comparable to a digital-twin~\cite{JonSNYH20} representation of a planned building. BEM has increasingly become a practical and supportive method for energy-efficient designs, operations, and retrofitting of buildings, with the aim of energy performance improvement and carbon emission reduction~\cite{PanZLYLYYJWZHHXYY23}. Fuelled by the rapid development of various data sensing, modeling, and visualizing technologies, BEM has attracted increased attention in application research for optimizing energy efficiency~\cite{WaibEC19}.

From a technical and definition point of view, BIM provides all the necessary data structures for BEM. However, BIM can hardly ever be directly used for BEM processes. BIM models rely on data input from various individuals and teams, including architects, engineers, and contractors. Human error during data entry can lead to inaccuracies and mistakes such as mislabeled components or missing data. Design changes in dynamic construction projects can occur frequently, and inconsistent updates in the BIM model can lead to errors. However, in the BEM process, very detailed and correct 3D information is required, as simulation models cannot account for uncertainties and errors. 3D models that can be used in a BEM process must be watertight. It further needs to be clear which building parts (e.g., windows, walls, rooms) are connected, and it needs to be defined which materials designers planned to use to construct the individual building parts.

As a result of this gap between BIM and the requirements for BEM, engineers increasingly started to invest time into developing automatic \emph{BIM-to-BEM conversion} processes. Converting a BIM model to a valid BEM model is a tedious 3D data wrangling process. It requires parsing the BIM information, running checks for various errors and inconsistencies that may occur, and providing algorithmic solutions for correcting these faulty definitions in the model. For example, building elements like walls or windows may be erroneously specified to overlap in BIM, or connections between rooms might be missing or placed wrongly. Ciccozzi~et~al~\cite{CicDPA23} analyzed various interoperability strategies, where the most published strategy is based on "standardized exchange formats and middleware corrective tools". These corrective tools try to reduce geometric errors but are tailored for specific use cases. Current planning tools do not offer widgets to guide users during conversion (e.g., point out errors and inconsistencies) and rely on manual data acquisition and maintenance. Automated algorithms~\cite{CicDPA23} for BIM-to-BEM conversion do not provide enough insights into the conversion process. Ultimately, HVAC engineers must rely on an automatic algorithm's BEM output for thermal calculations, not knowing which 3D BIM structures have been removed, added, or adapted.

\begin{figure}[t!]
    \centering
    \BeginAccSupp{method=pdfstringdef, ActualText={The image consists of three blocks. On top, "BIM Model" is written inside a box with black outlines. In the middle, a larger rectangle colored in red contains the five views: Relationship View, Element View, Space View, and BIM View. The collection of these views describes the BEMTrace workflow. At the bottom, "BEM Model" is written inside a box with black outlines. The "BIM Model" box and the "BEM Model" box are connected with the "BEMTrace" box in the middle via black arrows. The three boxes illustrate the data flow from a BIM model to a BEM model, with BEMTrace in the middle to understand the data transformation.}}
        \includegraphics[width=\columnwidth]{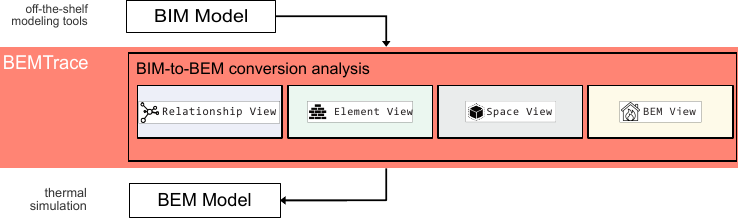}
    \EndAccSupp{}
    \caption{The BIM-to-BEM conversion process describes converting a BIM model to a BEM model. The novel contributions of this paper lie in the area highlighted in red.}
    \label{fig:workflowoverview}
\end{figure}

Working toward visual representations for BEM processes, we also had to deal with BIM-to-BEM data conversion, and our collaborators increasingly asked us for (visual) evidence concerning our BEM outputs. In this paper, we introduce BEMTrace, a visualization-driven approach for understanding the conversion from BIM to BEM models (Figure~\ref{fig:teaser}). Within the BIM-to-BEM conversion workflow, our proposed views and selection mechanisms enable users to understand the automated and so far black-box conversion process. Visualization proved to be an appropriate technique for increasing the engineers' understanding of the BIM-to-BEM conversion process. We contribute:
\begin{enumerate}
    \item \textbf{Introduction to BEM and BIM-to-BEM}: We introduce the application domain of Building Energy Modeling (BEM) and the transfer from BIM-to-BEM to the visualization community.
    \item \textbf{Exemplary use case for 3D data wrangling}: The BIM-to-BEM conversion process serves as a practical example where visualization is applied to support domain experts in understanding a tedious 3D data wrangling process.
    \item \textbf{Proposing context-adaptive selections}: We employ context-adaptive selections in our application that allow domain experts to follow building parts through the BEM conversion process.
    \item \textbf{Outlining directions for future work}: By introducing the application domains of BIM and BEM, and by outlining our application use case, we also identify ideas for future work and directions for visualization research in BIM, BEM, and building construction. 
\end{enumerate}

\section{Related Work}
\label{sec:relatedwork}

The work touches upon the topics of building information and energy modeling (BIM and BEM), BIM visualization and interaction, and visualization for data wrangling.

\paragraph{BIM and BEM} The central element in BIM is a 3D model of the planned construction, and the BIM standard enables the integration of additional information~\cite{HardMC15}. The construction process is dynamic, with many ad-hoc changes that must also be reflected in the BIM environment. The need for validation checks, especially for data exchange and interoperability, has, therefore, soon been realised~\cite{LeeES20}. Cann~et~al.~\cite{CannMPDJ22} developed a semi-automatic method for data quality assurance while working with BIM models. The successful implementation of BIM has led researchers to use this standard for further data analyses, employing AI~\cite{ZabGZA22} and ontology-based techniques~\cite{HerrCGM22}. In a BEM process, it is possible to simulate a building's energy demand under different environmental conditions~\cite{LiJ14} and by changing building parameters like heating and cooling capacity~\cite{RashKNLKL20} and the configuration of isolating layers~\cite{MoZh21}. In contrast to BIM, simulation models used in BEM are susceptible to errors and inconsistencies in the 3D building model. Thermal simulations require a well-defined database with a clear definition of spatial arrangements (i.e., building part positions and connections between parts) and used building parameters (i.e., materials and layers). Therefore, thermal simulations in BEM can hardly ever be run directly on BIM data~\cite{MailDBR13}. Converting a BIM model to a BEM model requires detailed checks for various errors and inconsistencies and provides algorithmic solutions for correcting these faulty definitions in the model~\cite{RamMM20}. Some of the existing BIM-to-BEM conversions are targeted toward specific use cases to limit the number of possibly necessary operations~\cite{CheDCC23}. Kiavarz~et~al.~\cite{KiavJRS23} built a graph-based BIM-to-BEM algorithm for more general applications and for different use cases. In contrast to Kiavarz~et~al., we do not create new geometry but modify BIM parts. \textbf{We introduce a visualization application to understand BIM-to-BEM conversions, which is an underrepresented topic in visualization research (Contribution 1 \& 4).}

\paragraph{Data Wrangling} The process of data wrangling describes transforming raw data into a structured format suitable for analysis~\cite{NiraSDK23}. This often includes tasks such as data cleaning, handling missing or inconsistent data, formatting data types, removing duplicates, and merging datasets~\cite{PatH18}. Data wrangling is considered an essential first step in a multi-stage data science process~\cite{KandelPHH12}, which is time-consuming but is also an essential step for analysts to be able to rely on high quality data afterwards~\cite{FernKKPPS23}. Visualization researchers have acknowledged the importance of supporting data analysts in data wrangling~\cite{KandeHPKVRWLBB11}. Data scientists leverage a tight coupling between visualization and data transformation code~\cite{PuK23} regularly. Visual representations aid in data-focused and tedious tasks such as data transformations and promote clarity and transparency~\cite{MilLPM21}. Visualization researchers proposed visualization approaches for data wrangling that target tabular data~\cite{KandelPHH11}, text and news data~\cite{KasBM23}, and time series data~\cite{BernHRPBK19}. Mühlbacher~et~al.~\cite{MuePGSS14} showed how visual data representations enable user involvement in long-running, ongoing algorithmic computations. Other techniques depicted data transformation steps, similar to provenance visualization. Visualizing the data wrangling steps can either be done by visualizing directly the used data wrangling scripts~\cite{XioFDLYCBW23} or by concentrating on the semantics of the scripts~\cite{XioLFWXW23}. On a more abstract level, provenance-based techniques help to store operations on data~\cite{RaganESC16}. Interactive, provenance-based applications like \emph{DQProv Explorer}~\cite{BorsGM19} help users keep track of the data wrangling steps executed in the preprocessing phase. DQProv Explorer employs a provenance graph to store the data wrangling operations and provides interactive visualizations that help analysts to keep track of the data-quality development over time. In provenance visualization research, Gratzl~et~al.~\cite{GratzLGCS16} proposed \emph{CLUE}, a model for reproducing, annotating, and presenting visualization-driven data exploration. CLUE allows users to interactively switch between a presentation stage (focused on insights) and an exploration stage (focused on interactive data exploration). This way, users can keep track of their paths/thoughts/ideas to come to certain insights and switch from exploration to storytelling and back. Visualization solutions for data wrangling like DQProv Explorer~\cite{BorsGM19} do not cover 3D data yet. Provenance approaches like CLUE~\cite{GratzLGCS16} focus on keeping track of insights, but not the raw data. \textbf{We outline an exemplary use case of using data visualization to understand a complex 3D data wrangling process (Contribution 2).}

\paragraph{BIM Visualization and interaction} Visualization techniques for displaying 3D models are already well established in several market-ready applications~\cite{TeyC09}. Rendering and exploring 3D models is essential to state-of-the-art computer-aided design (CAD) applications used in construction~\cite{Encar87}. CAD models often serve as a fundamental data structure in different applications, such as visualization solutions for smart manufacturing~\cite{ZhoLLZXRXR19} and Visual Analytics for urban environments~\cite{DenWLTXW22}. CAD models are also often employed for Extended Reality (XR) and immersive analytics applications~\cite{HurtMGCR22}. The visualization of BIM models is a relatively new topic in visualization research. As one of the first challenges, Johansson~et~al.~\cite{JohRBS15} dealt with the multitude of information in BIM in real time. Especially since 2012, the number of visualization solutions for BIM has steadily increased~\cite{IvsMQSC20}. Most of these visualization applications concentrate on the design and operation phase and combine scientific and information visualization techniques~\cite{WuCCLWY19}. The complexity of BIM visualizations steadily increased over time and nowadays also allows the user to represent BIM models in a web browser~\cite{WanHSYB23}. Alongside 3D rendering, advanced techniques for interaction and selection in 3D have been developed~\cite{HagEVS01}. With 3D data becoming more complex, researchers concentrated on simplifying interactions and selections based on the current rendering context. \emph{Context-aware selection} techniques can deduce a user's selection intent based on the underlying data and view conditions~\cite{YuEII12}. They have already been successfully applied to volume data~\cite{WiebVFH12}, volumetric structures~\cite{KohBKG09}, point clouds~\cite{YuEII16}, and dense graphs~\cite{HoraBSDT21}. Context-aware selections have also been proven to be useful in immersive Virtual Reality (VR) settings~\cite{ZhaIXLY24}. Such selections enable users to directly interact with the rendered objects currently seen on screen or in the immersive environment. Researchers also concentrated on adaptive rendering techniques, where the 3D renderings are adjusted based on the currently visible data. Miao~et~al.~\cite{MiaDSAKIGBV18} proposed a multi-scale visualization approach for the representation of DNA nanostructures, where visualizations and interactions are adapted according to the current semantic abstraction level selected by the user. Smooth transitions between the different renderings create a continuous, multiscale visualization and interaction space. \emph{Barrio}~\cite{TroidCGPHB22} follows a similar idea, where visualizations are automatically adjusted based on user selections and the number of structures to compare. The system switches between small multiples of spatial 3D views, abstract quantitative views, and arrangements like linked and juxtaposed views. \textbf{We propose \emph{context-adaptive selections} (Contribution 3), a novel interaction technique that allows users to switch between different contexts in our application. Context-adaptive selections combine the ideas of context-aware selections~\cite{YuEII12} and adaptive rendering~\cite{TroidCGPHB22}.}

\section{Introduction to BIM and BEM}
\label{sec:background}

\textbf{Building Information Modeling (BIM)} defines a digital, multi-dimensional platform for managing a digital representation of a building or infrastructure project before and while it is built~\cite{Casi22}. BIMs serve as comprehensive information repositories, encompassing geometric and spatial data and detailed specifications, quantities, and other attributes. This holistic approach allows stakeholders from various disciplines, such as architects, engineers, contractors, and facility managers, to collaborate more effectively and make informed decisions at every project stage. Traditional building design primarily relied upon two-dimensional technical drawings (plans, elevations, sections, etc.). Building Information Modeling extends construction processes with 3D models, incorporating time, costs, asset management, sustainability, and more information. BIM covers geospatial information, quantities, and properties of building components and enables a wide range of collaborative processes relating to the built asset.

The ISO standard no. 19650-1:2018~\cite{ISO18} defines BIM as '\emph{use of a shared digital representation of a built asset to facilitate design, construction and operation processes to form a reliable basis for decisions}'. It is worth noting that BIM also arose from poor software interoperability in construction, which has long been regarded as an obstacle to industry efficiency in general. To achieve interoperability between applications, neutral, non-proprietary, or open standards for sharing BIM data among different software applications have been developed. The open exchange format \emph{Industry Foundation Classes (IFC)} empowers interdisciplinary cooperation and reduces update intervals between partners. The IFC model is object-oriented and incorporates 3D geometry, spatial relationships (topology), quantities, and building part properties (semantic).

\begin{figure}[b!]
    \centering
    \BeginAccSupp{method=pdfstringdef, ActualText={This image contains two 3D renderings of a building floor. Only the walls are rendered. The walls are colored according to simulation data, where the room temperature has been calculated. In the rendering on the left, all walls are red (apart from a thin stripe toward the ceiling). In the rendering on the right, all walls are yellow to orange. The images depict the difference in room temperatures without cooling (left side) and with cooling (right side) in place.}}
        \includegraphics[width=\columnwidth]{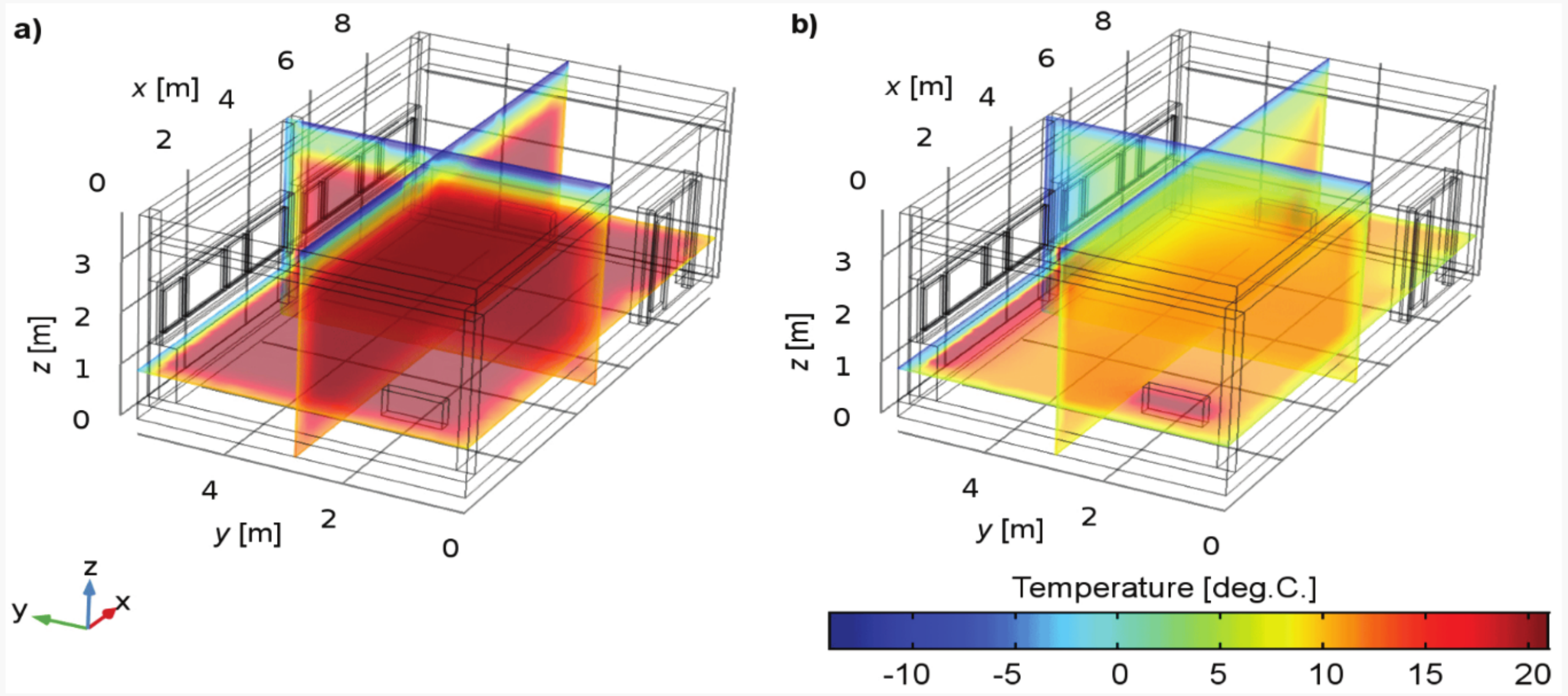}
    \EndAccSupp{}
    \caption{Building Energy Modeling (BEM) allows engineers to run complex physical simulations to study a room's temperature under different conditions. In this example, temperatures without (a) and with (b) cooling in place are compared. Image taken from Charvátová~et~al.~\cite{CharPZ18}.}
    \label{fig:bem}
\end{figure}

For the practitioners involved in a project, BIM enables a virtual information model where each professional adds only discipline-specific data to the shared model. BIMs are typically organized into various layers or levels of detail, where each layer represents different aspects of the project. The structural level, for example, focuses on the building's structural components (i.e., beams, columns, foundations) while the mechanical layer comprises heating, ventilation, and air conditioning systems. Stakeholders are granted access to the relevant layers based on their roles and responsibilities in the project. As such, BIM offers a number of advantages, including improved communication and coordination among project teams, reduced errors and clashes, and enhanced visualization and simulation capabilities. 

Thermal simulations~\cite{DAmbrPal20} represent a specific subcategory of computational modeling, wherein the objective is to analyze the thermal behavior and heat transfer within physical systems.
These simulations employ mathematical and computational techniques to simulate how heat flows and interact with materials, components, and environments. Thermal simulations help to design and optimize HVAC (heating, ventilation, and air conditioning) systems for efficient heat management, to ensure thermal comfort, and to address various thermal-related challenges in complex systems.

\textbf{Building Energy Modeling (BEM)} is the practice of creating a digital representation of a building's physical and operational characteristics~\cite{JohPSZW20}. Figure~\ref{fig:bem} shows an example of simulation results created in BEM. BEM combines the information of BIM with thermal simulation to assess and optimize the building's energy consumption, heating, cooling, lighting, and other systems to improve energy efficiency and sustainability. BEM allows architects, engineers, and energy professionals to make informed decisions about building design, HVAC systems, insulation, lighting, and other factors. Testing different parameters in BEM can help designers reduce energy consumption, lower operational costs, and minimize environmental impact.

BIM and BEM focus on different tasks. BIM facilitates the construction process and provides an interoperable tool for all stakeholders. BEM focuses on energy simulation and heating, ventilation, and air conditioning system prediction. The interplay of all thermal exchanges requires a detailed description of how rooms are connected. Although BIM provides the basis for storing this information, users do not necessarily correctly enter the connections between building parts into the model. In the course of the modeling process, users must follow the vendor's specific guidelines to generate high-quality data sets. To improve interoperability between BIM tools, exchange formats for specialized use cases have been implemented. Generalized certification has not yet been realized to its full potential, so relevant details from previous BIM tools may get lost in the course of exchange processes.

\begin{figure}[t!]
    \begin{tikzpicture}[outer sep=0, inner sep=0, font=\sffamily\small]
        \BeginAccSupp{method=pdfstringdef, ActualText={Three renderings next to each other depict how a wall element was changed in the conversion process. In the first image, the raw data shows a wall (in green) and a window (in red) in the middle. The following image shows the same wall with a hole where the window was before. In the last image, the wall surface is enlarged to touch the other walls next to it.}}
            \node[anchor=north west] at (0,0) {\includegraphics[width=.98\columnwidth]
            {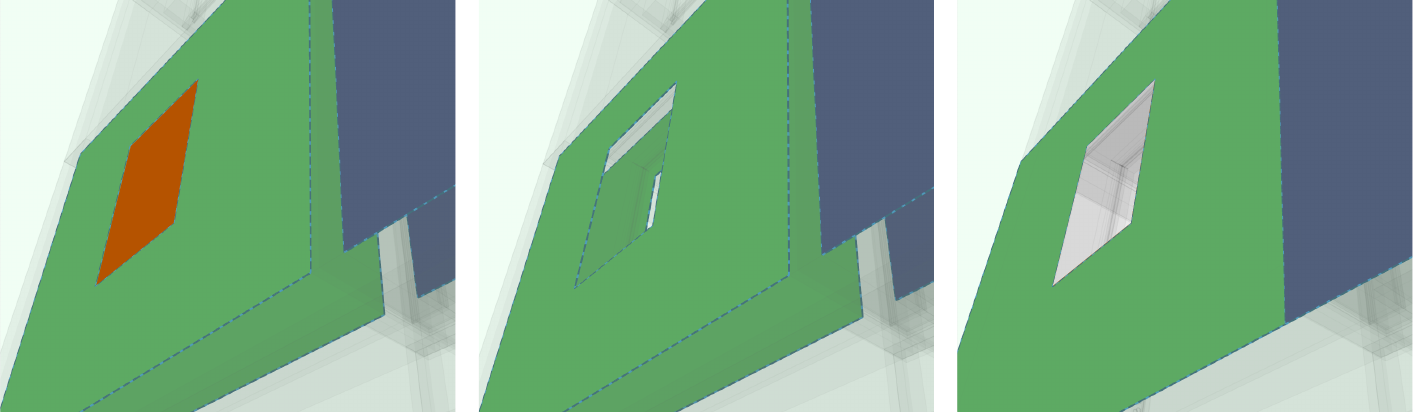}};
            \node at (1.5,-2.8) {Raw data};
            \node at (4.4,-2.8) {Removing recess};
            \node at (7.3,-2.8) {Enlarging surfaces};
        \EndAccSupp{}
    \end{tikzpicture}
    \caption{Thermal simulations in BEM rely on watertight arrangements of space boundaries. Since surfaces in BIM are not always watertight, data preprocessing involves removing recesses and appropriately enlarging surfaces between rooms.}
    \label{fig:bimtobem}
\end{figure}

Due to this mismatch between BIM and BEM needs and structures, automated methods for \textbf{BIM-to-BEM} conversion have been developed~\cite{KiavJRS23}. Conversion methods usually rely on graph-based structures to represent connections between building parts. The building parts depicted in BIM are geometrically set into relation and placed into a spatial graph structure. Missing parts (e.g., connecting layers) are inserted if needed. An example of necessary adjustments during BIM-to-BEM conversions is shown in Figure~\ref{fig:bimtobem}. Ultimately, a BEM model consists of BIM objects with additional information about which parts are connected.

\section{Thermal Simulation Use Case}
\label{sec:usecasedescr}

Our use case stems from the need to provide BEM computations to our project partners. We started with a BIM model and, similar to previous approaches~\cite{PorsDSF21}, very soon realized the need for BIM-to-BEM conversion before being able to start thermal simulations.

\subsection{Data}

Our project partner used \emph{Archicad} by Graphisoft~\cite{Graph24} to construct example buildings for further calculations, exporting them as IFC files. The current focus of our domain experts is on private housing, less on analyzing office and commercial buildings. The BIM models used in the study mainly follow the structure of general housing environments in Europe, with mostly rectangular rooms and airspace around the buildings. We use existing libraries to load BIM data from IFC files.

\subsection{Terms}

\begin{figure}[t!]
    \begin{tikzpicture}[outer sep=0, inner sep=0, font=\sffamily\small]
        \BeginAccSupp{method=pdfstringdef, ActualText={One image on the left shows a building floor, rendered in 3D, where two rooms are connected by walls that form a T. Divided by a middle line, two more images are shown on the right. In these two images, space boundaries are marked in red. In the first one, space boundaries of type A, the large areas on the wall that connect two rooms, are marked in red. In the second one, space boundaries of type B, the thin stripe where the T-wall connects with the first room, are marked in red.}}
            \node[anchor=north west] at (0,0) {\includegraphics[width=.98\columnwidth]{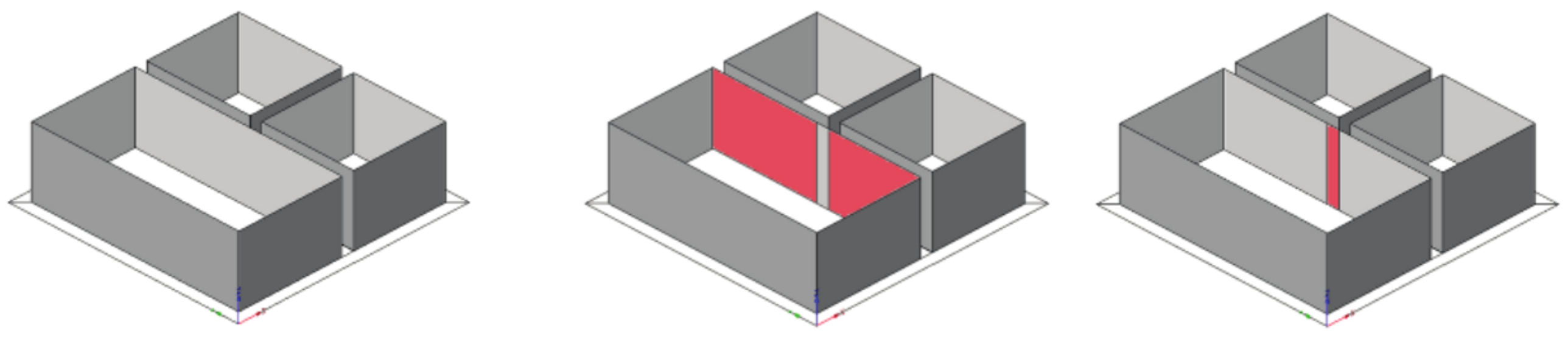}};
            \draw (2.9,0) -- (2.9, -2.2);
            \node at (1.35,-2.1) {Wall};
            \node at (5.9,-2.1) {Space boundaries (type A and B)};
        \EndAccSupp{}
    \end{tikzpicture}
    \caption{Space boundaries define connecting surfaces between two rooms. A wall element may consist of more than one space boundary, as shown in this illustration. Figure parts taken from buildingSMART~\cite{bs24}.}
    \label{fig:spaceboundaries}
\end{figure}

A comprehensive list of terminology used in the BIM-to-BEM conversion is given in Table~\ref{tab:terms}. We concentrate on the BIM concepts of building \emph{elements} and \emph{spaces}. Elements are physical parts of a building, including walls and windows. Spaces are geometric volumes whose dimensions are defined by the surrounding elements (e.g., walls defining a room). Spaces may reside inside or outside a building, where spaces outside represent the air or earth surrounding the building. The planar shape where an element and a space touch is called a \emph{space boundary}. Elements need at least one but may have several space boundaries attached. For example, the middle wall in Figure~\ref{fig:spaceboundaries} has several space boundaries attached as it is connected to more than one space.

As an ultimate goal in the BIM-to-BEM conversion, we assemble a topologically sound neighborhood graph of elements, spaces, and space boundaries. We compute \emph{connections}, which establish a link between two spaces defined by space boundaries. A connection may be, for example, created by a door between two spaces. 
In the context of BEM, connections represent a fundamental aspect of thermal energy exchange. The material types of the elements along a connection define how thermal energy can be transferred. For example, external doors are usually made of different materials than internal ones. The thermal energy exchanged via the connection created by an internal door will differ from the thermal energy transported via an external door. A space, together with all attached connections to other spaces, is called a \emph{room} in BEM. A \emph{room network} summarizes all rooms and their connections and is used as input for thermal simulations.

\begin{table}[h]
    \centering
    \caption{Common terminology used in the BIM-to-BEM conversion workflow and in this paper.}
    \label{tab:terms}
    \BeginAccSupp{method=pdfstringdef, ActualText={The table has three columns, and rows are interchangeably displayed in white and grey.}}
        \begin{tabular}{@{}lll@{}}
            \toprule
            \textbf{World} & \textbf{Term}    & \textbf{Meaning}
            \\ \midrule
            BIM & Element & \begin{tabular}[c]{@{}l@{}}A physical object of a building\\ (e.g., wall, door, slab, window).\end{tabular} \\
            \rowcolor[HTML]{D8D8D8} 
            BIM & Space & \begin{tabular}[c]{@{}l@{}}A geometric volume classified as\\ internal or external (air, earth).\end{tabular} \\
            BIM & Space Boundary   & \begin{tabular}[c]{@{}l@{}}The touching plane between one\\space and one element.\end{tabular} \\
            \rowcolor[HTML]{D8D8D8} 
            BEM & Connection & \begin{tabular}[c]{@{}l@{}}An abstract connection \\ between two spaces, defined by\\ two space boundaries.\end{tabular} \\
            BEM & Room & \begin{tabular}[c]{@{}l@{}}Represents one space and all\\its outgoing connections.\end{tabular} \\
            \rowcolor[HTML]{D8D8D8} 
            BEM & Room Network & \begin{tabular}[c]{@{}l@{}}A collection of several rooms that\\serves as input for a thermal\\simulation.\end{tabular} \\
            \bottomrule
        \end{tabular}
    \EndAccSupp{}
\end{table}

\subsection{Task Analysis and Goals}

Converting a BIM model to a BEM model is a tedious process that requires complex geometric data checks and adjustments (Section~\ref{sec:background}). We work together with two civil engineers who deal with the design, construction, and maintenance of buildings, mostly private housing. These domain experts have a thorough knowledge in building physics and building energy computations. They are unfamiliar with geometric computer graphics operations, like cutting out recesses (Figure~\ref{fig:bimtobem}) or finding surface pairs based on the normal vectors. The civil engineers asked us to explain the BIM-to-BEM conversion process visually. The domain experts required to understand how data in BIM is used in the BIM-to-BEM conversion process. They needed to be able to trace BEM information (e.g., connections, rooms) back to raw data in BIM (e.g., elements, space boundaries). Based on these requirements, we identified the following tasks:

\begin{itemize}

    \item \textbf{T1 - Loading a BIM model and checking the spatial arrangement of elements:} The initial setup, and at the same time, input data for the conversion process, is based on the BIM model. Users must load, view, and explore this model to check the geometric arrangement of elements in 3D and use visualization functionality to identify possible faults.

    \item \textbf{T2 - Inspecting data quality issues that hinder an automatic conversion:} We automatically detect 3D geometry conflicts in the BIM model that may cause problems during the conversion process. Users can analyze the automated resolution of these conflicts and update the BIM model or the conversion options, if necessary.
    
    \item \textbf{T3 - Converting a BIM model to a BEM Model:} Users must be capable of initiating the automatic BIM-to-BEM algorithm.
    
    \item \textbf{T4 - BEM model visualization and inspection:} The BIM-to-BEM result is a BEM model suitable as input for thermal simulations. Users need to be able to view this model in 3D and study the rooms and their spatial arrangement. Users must analyze individual aspects of the BEM model in more detail and filter the complete model.
    
    \item \textbf{T5 - Tracing BEM information to the raw data stored in BIM:} Users need to understand how individual parts in BEM (e.g., rooms, connections) have been computed, and if space boundaries were changed during the BIM-to-BEM conversion.
    
\end{itemize}

The tasks led to the specification of the following goals:

\begin{itemize}
    \item \textbf{G1 - 3D overview of the BIM model:} Provide users with an interactive 3D spatial view on the input BIM data. This also includes possible geometry conflicts that might influence the data conversion process.
    \item \textbf{G2 - BIM-to-BEM conversion:} Deliver an automatic conversion from BIM-to-BEM.
    \item \textbf{G3 - 3D overview of the BEM model:} Provide users with an interactive 3D spatial view of the BEM model computed by the BIM-to-BEM conversion.
    \item \textbf{G4 - Insights into the data conversion:} Increase users' trust in the automatic BIM-to-BEM conversion by enabling visual, interactive views of the different computing steps.
\end{itemize}

\begin{figure}[b!]
    \centering
    \BeginAccSupp{method=pdfstringdef, ActualText={In the first row, a bulding is shown, which is then disassemble into two storeys. In the second row, the storeys are further disassembled into building elements, shown as walls and windows, spaces (rooms), and space boundaries (connections).}}
        \includegraphics[width=\columnwidth]{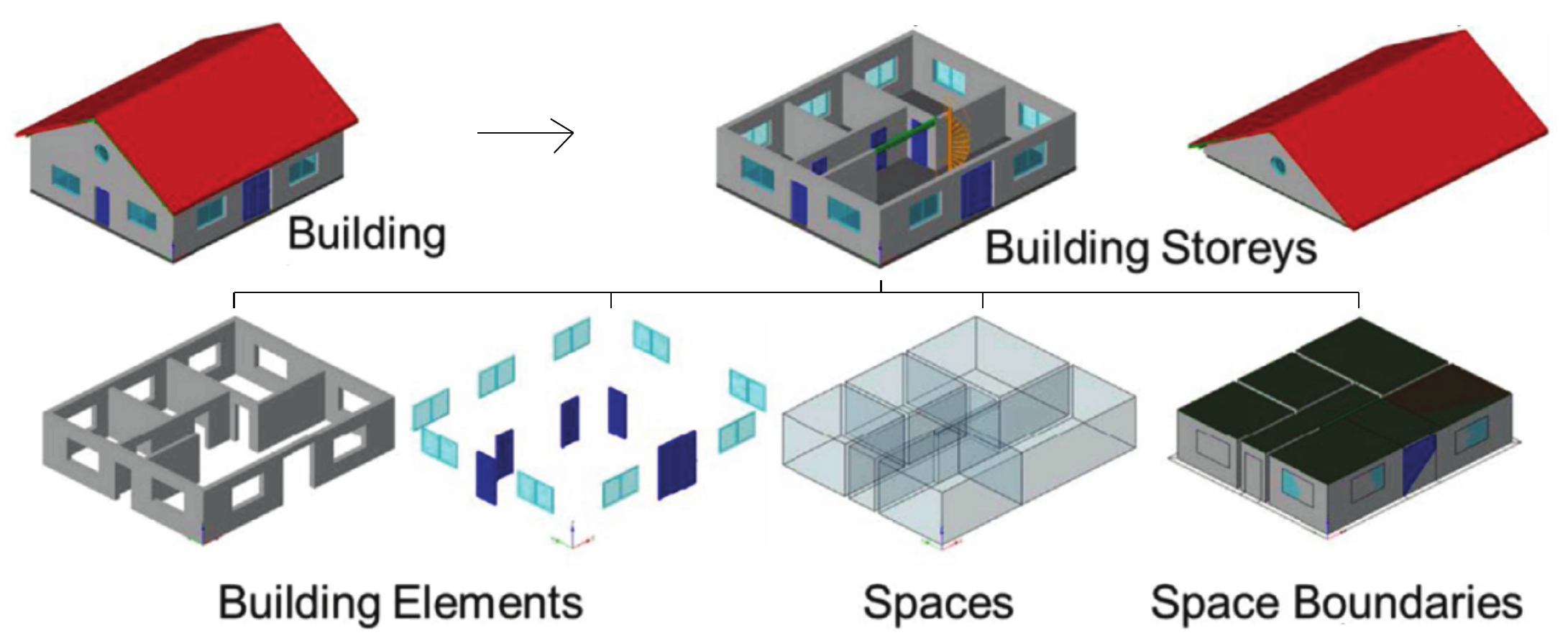}
    \EndAccSupp{}
    \caption{In BIM, a \emph{Building} is composed of \emph{Building Storeys}, which are further composed of \emph{Building Elements}, \emph{Spaces}, and \emph{Space Boundaries}. We follow this notion and division of BIM structures in our application. Images taken from Malhotra~et~al.~\cite{MalhBNHESAFVOS22}.}
    \label{fig:ifc}
\end{figure}

\section{BEMTrace Application}
\label{sec:bemtrace}

In close collaboration with the domain experts, as suggested in the Nested Model~\cite{Munz14}, by using rapid prototyping and several evaluation sessions, we developed the BEMTrace approach, which supports the tasks and goals defined in Section~\ref{sec:usecasedescr}. BEMTrace consists of five views of the BIM and BEM data to help the users understand how we create a BEM model from a BIM model. Users can now interactively explore the data and trace information they see in BEM back to the raw data stored in BIM.

\subsection{BEMTrace Parts}

BEMTrace consists of two parts, in the following called \emph{worlds}. One part represents the \emph{BIM World}, and the other one represents the \emph{BEM World}. The worlds represent the two context areas domain experts usually deal with. In both worlds, we provide different data visualization views showing various aspects. Figure~\ref{fig:teaser} gives an overview of the BIM World and the BEM World and their connected views. By assigning views to different worlds, we connect them with different types of data (i.e., raw data in BIM or generated results in BEM). The BIM World covers visualizing the input data from different perspectives (spaces and elements). On the way of converting the BIM model to a BEM model, relationships and conflicts can be studied in the so-called Relationship View. The BEM World represents the final BEM model computed by the BIM-to-BEM conversion.

\subsubsection{BIM World}

In the BIM World, users load and visualize BIM models (G1). The three views we use (BIM View, Space View, Element View) are described in the following and illustrated in Figure~\ref{fig:teaser}. A schematic illustration of the information contained in a BIM model is shown in Figure~\ref{fig:ifc}. We use the existing classification of BIM parts to design the three views in the BIM World since the domain experts were already familiar with it (i.e., seeing the building as a whole and inspecting the individual parts of elements and spaces).

Users can view the BIM 3D model and select individual parts to view more detailed information (T1) in the \IconBIMView \textbf{BIM View}. 
Traditional rendering is enriched with the ability to interactively isolate regions of interest. A low-opacity silhouette of the rest of the building is rendered to preserve context. We use state-of-the-art 3D rendering techniques and also enable a photorealistic representation of windows and doors.

In the \IconSpaceView \textbf{Space View}, we give an overview of a building's topological organization. Every space has a specific geometry that defines its volume. Outside regions around the building (air, earth) are also interpreted as spaces. This information is later needed in BEM to calculate thermal energy transport between the building and its exterior environment. In the Space View, we omit the visualization of the related elements. The space color helps users distinguish between internal spaces (\textcolor{myblue}{blue}), air spaces (\textcolor{mygreen}{green}), and earth spaces (\textcolor{mybrown}{brown}). In the Space View, users can check if automatic heuristics correctly assigned space types and if the size and location of spaces look plausible.

In the \IconElementView \textbf{Element View}, we represent the building's elements. We only show elements and omit information about spaces. Users can check the size and spatial arrangement of elements, the elements' geometrical orientation, and the elements' material properties in this view. The visualization comprises a 3D representation of all elements, while a shared color emphasizes classifications of elements with the same types of material composition.

\begin{figure}[b!]
    \begin{tikzpicture}[outer sep=0, inner sep=0]
        \BeginAccSupp{method=pdfstringdef, ActualText={The image shows the entire GUI of the BEMTrace application. On top, we see a menu bar that allows users to select different views. Below, on the left, we see a 3D view, which currently shows a 3D rendering of a building. The space boundaries are rendered as colored planes. On the right the menu controls and GUI elements are shown, in dark mode.}}
            \node[anchor=north west] at (0,0) {\includegraphics[width=.98\columnwidth]{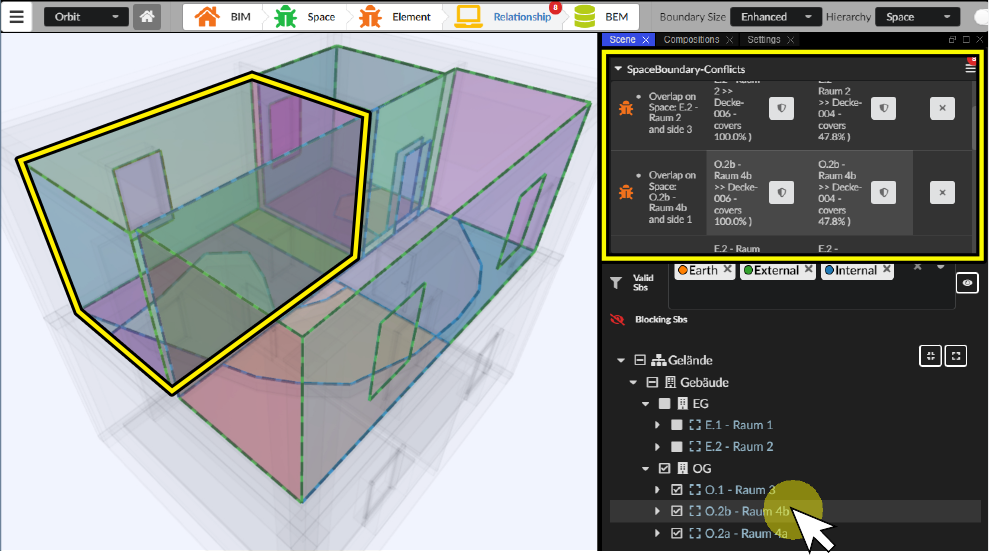}};
            \node[anchor=north west] at (3.2,-5) {\includegraphics[height=0.4cm]{pics/icon-relationshipview}};
        \EndAccSupp{}
    \end{tikzpicture}
    \caption{BEM connections are created based on space boundaries. In the Relationship View, users inspect these space boundaries (left side, 3D view). Additionally, users are informed about data errors (right side) and are able to examine them.}
    \label{fig:relationshipview}
\end{figure}

\begin{figure*}
    \begin{tikzpicture}[outer sep=0, inner sep=0, font=\sffamily\small]
        \BeginAccSupp{method=pdfstringdef, ActualText={We see four 3D views: BEM View on the left, next to it the Relationship View, then the Element View, and on the right the Space View. The views show a building rendered in 3D. In the BEM View, a 3D rendering is shown, with a wall selected and outlined with red lines. In the Relationship View, the two rooms connected to this wall are shown in color, with the rest of the building in a ghosted shape. In the Element View, the wall is rendered solid and colored blue. In the Space View, the two connected rooms are rendered as colored blocks (green for the outside room and blue for the inside room).}}
            \node[anchor=north west] at (0,0) {\includegraphics[width=\textwidth]{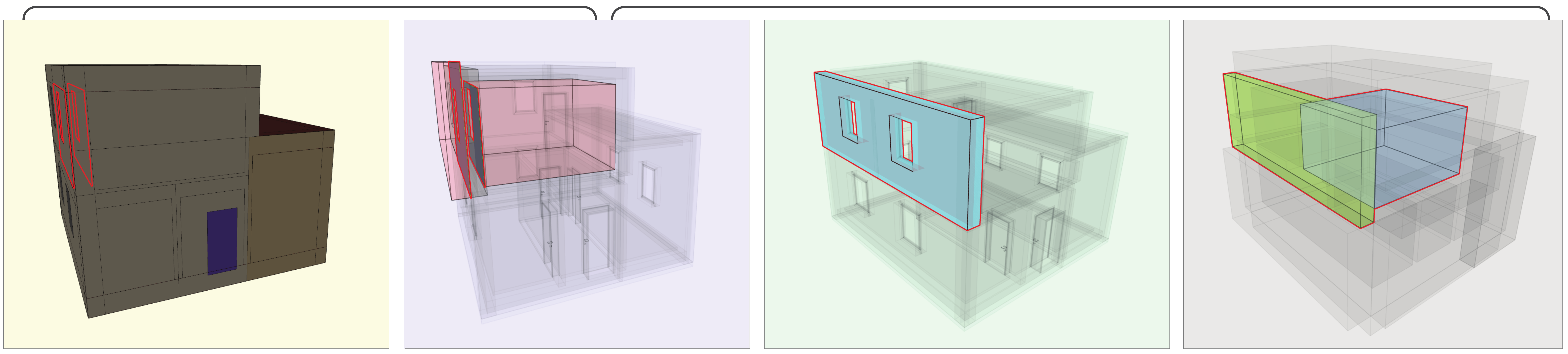}};
            \node at (3.5,.1) {BEM World};
            \node at (12.5,.1) {BIM World};
            \node[anchor=north west] at (1.6,-4.11) {\includegraphics[height=0.4cm]{pics/icon-bemview}};
            \node[anchor=north west] at (5.3,-4.11) {\includegraphics[height=0.4cm]{pics/icon-relationshipview}};
            \node[anchor=north west] at (10,-4.11) {\includegraphics[height=0.4cm]{pics/icon-elementview}};
            \node[anchor=north west] at (15,-4.11) {\includegraphics[height=0.4cm]{pics/icon-spaceview}};
        \EndAccSupp{}
    \end{tikzpicture}
    \caption{Context-adaptive selections are an interaction mechanism in BEMTrace to track rooms and connections in BEM back to their original raw data (space boundaries, elements, spaces) in BIM. Users start by selecting a room or a connection in the BEM View. In the other views, only the information fitting the context is shown. In the Relationship View, the space boundaries can be inspected. In the Element View, the elements comprising the space boundaries are highlighted. In the Space View, the spaces connected by space boundaries are shown.}
    \label{fig:selections}
\end{figure*}

\subsubsection{Relationships and Conflicts}

On the way from BIM to BEM, between the BIM and the BEM World, we included the \IconRelationshipView \textbf{Relationship View}. The Relationship View covers data and information from both the BIM and the BEM World. The view uses the 3D information of the BIM model as a basis to display possible geometric conflicts hindering a successful BIM-to-BEM conversion. Only in the Relationship View, users can inspect the space boundaries of a model in 3D, as illustrated in Figure~\ref{fig:relationshipview}. Connections in BEM are defined by space boundaries originating from the same element. This makes space boundaries an essential element in correctly computing a room network for BEM. In the 3D view, users can see the space boundaries and study their spatial arrangement. Space boundaries and their types are encoded in color.

An essential function of the Relationship View is to provide information about possible geometry conflicts that might hinder the BEM conversion process (T2). We automatically detect overlaps of spaces, elements, and space boundaries in the BIM model and identify them as conflicts. Conflicts occur due to duplicate and overlapping space boundaries, space boundaries that are wrongly placed or created (i.e., geometrically impossible connections), space boundaries that are too small, or space boundaries that have been placed inside elements. These geometry conflicts are of special interest in BEM. In the example of Figure~\ref{fig:relationshipview}, on the right side, the user is informed about conflicts caused by overlapping space boundaries. Conflicts are presented in a list view, where a sort mechanism groups semantically connected conflicts to convey a quick overview and interaction possibility. When users click on a list item, more information about the conflict is displayed. In the 3D view, we draw the space boundaries connected to a conflict, alongside with the related spaces and elements. Users can directly resolve conflicts in the Relationship View, leading to BIM parts being ignored in the conversion process (visualizations will be re-rendered). Users can also adjust the BIM model outside BEMTrace and reload it.

All rooms must be fully delimited by connections (i.e., being watertight) to ensure energy conservation in the simulation. As mentioned in Section~\ref{sec:background}, the recesses of windows and doors need to be removed from the space boundaries, and uncovered areas to neighboring space boundaries must be covered to ensure that the thermal flow between rooms is physically reliable. Adjustments are made automatically based on the BIM model. In the Relationship View, users can visually inspect the differences and are informed about cases where the conversion fails.

\subsubsection{BIM-to-BEM Conversion}

Users launch the BIM-to-BEM conversion (G2, T3) based on the model that is currently loaded. We perform initial sanity checks where we inspect all spaces and element classifications. In the case of duplicate information, the conversion might ignore spaces or elements and the attached space boundaries. Invalid space boundaries or space boundaries without a matching partner are also ignored.

The goal of the BIM-to-BEM conversion is to build a room network where nodes represent rooms and edges represent connections that relate to thermal flow. Connections are specified by two paired space boundaries attached to the same element. The edges are categorized by the involved spaces (e.g., internal, air, earth), the elements (e.g., wall, slab, window), and the elements' material compositions. The connection area depends on the dimension of the underlying space boundaries. Connections build a graph. After the BIM-to-BEM conversion, all rooms are connected via edges. For realizing the BIM-to-BEM conversion, we did not implement a novel solution but followed the graph-based conversion proposed by Kiavarz~et~al.~\cite{KiavJRS23}. While Kiavarz~et~al. extract geometry features from meshes, we rely on existing space boundaries for computing connections.

\subsubsection{BEM World}

After the BIM-to-BEM conversion, users can explore the computed rooms and connections that form the BEM model (G3, T4). We currently provide one \IconBEMView \textbf{BEM View} in the BEM World. Users can concentrate on individual rooms and connections by filtering. For example, users might want to see connections of a certain type (e.g., created by windows or doors), or assigned to a specific material composition. Filtered rooms and connections are highlighted, while low-opacity silhouettes of the remaining BEM model structures provide context. An alpha parameter controls the transparency in the rendering and can be manually adjusted.

\subsubsection{Interactions}

Since all our views depict 3D information, we provide users with familiar 3D interactions, such as zooming, panning, selection, and hovering. Users are able to explore the original BIM and the derived BEM models interactively. The 3D views are augmented with transparency and outline rendering to improve spatial awareness in the presence of occlusions. Additionally, users may interactively focus on specific, view-dependent categories using a tree view. For example, categories might include the lower and upper floor in the BIM View, or connections according to material composition in the BEM View. A screenshot of the user interface with the 3D view and tree view next to each other is shown in Figure~\ref{fig:relationshipview}.

\subsection{Context-Adaptive Selections}

The novelty in our proposed workflow lies in the users' ability to interactively switch between all views and the adaptivity of the user interface based thereon. By selecting individual rooms or connections in the BEM View, users can explore how this information was created based on BIM data (G4, T5). To keep selections and filters consistent across all worlds and views, we introduce \emph{context-adaptive selections} as an interaction mechanism.

Users start an exploratory analysis by selecting a room or connection in the BEM View either in the 3D view or in the tree view. The selected room or connection is then marked in red in the rendering and highlighted in the tree view. By clicking on the other views, users can trace the selected object step-by-step through all computation stages of the BIM-to-BEM conversion. Figure~\ref{fig:selections} illustrates a selection process where the users picked a connection. First, users inspect the space boundaries in the Relationship View. Here, they see the location and size of the space boundaries, and they can observe and compare how the conversion algorithm changed the space boundaries (i.e., a recess window was cut out). Switching to the Element View, users inspect the element (i.e., wall) that is connected to the space boundaries. Switching to the Space View, the two spaces adjacent to the space boundary are shown. It can be seen that the wall connects an internal with an external space, thus representing an external wall.

Users can only select objects attached to a view. In the BEM View, for example, users can select either rooms or connections. In the Space View, users can select spaces. By letting users interact only with objects that are available in the current view, we follow the idea of context-aware selections~\cite{YuEII12}. Users can select objects and then switch to another view. However, a selected object, e.g., a wall, will not be available in all views. For example, the Space View does not support elements like walls. Selected objects are, therefore, only shown if they are available.

In cases where the currently selected object has no representation within the view (e.g., walls not available in the Space View), context information related to the selection is highlighted. For example, the spaces connected to a selected wall are highlighted in the Space View.
We here follow the idea of adaptive rendering~\cite{TroidCGPHB22}. Combining both approaches led to our proposed concept of \emph{context-adaptive selections}. Context-adaptive selections allow users to study a phenomenon (i.e., a connection or an element) from different perspectives in different views. Context-adaptive selections serve the following purposes in the BEMTrace workflow:
\begin{itemize}
    \item \textbf{Selections based on current context} Users can only operate on building objects that are available in the current world and view.
    \item \textbf{Context-specific information} The different views provide information about different aspects of the currently selected objects.
\end{itemize}

If the user selects an object in one of the views, we store the information about the selected object. In the other views, we perform queries to find out which parts should be highlighted in the view currently activated by the user. For example, if the user selects a connection in the BEM View, we query for the element that is related to this connection in the Element View, and for the connected spaces in the Space View.

Selections also work in the other direction, i.e., from BIM data to the BEM View. Starting from BEM, users can understand how a certain connection was created from the raw data. Starting from BIM or any other view, users can inspect how specific objects have been integrated during the conversion process.

\subsection{Implementation}

Our application builds upon the \emph{Aardvark}~\cite{Aardvark24} platform, which is mainly written in F\#. We employ \emph{Aardvark}'s rendering engine~\cite{HaasSMT15} to display the 3D views. The user interface and data loading functionality have been implemented using Open Source software. The user interface is based on web-technologies (HTML and JavaScript). We apply the \emph{GoldenLayout} library~\cite{Deepstr24} for multi-view window ability and the \emph{Fomantic} library~\cite{Fomant24} for stylizing. We use the .NET library \emph{xBIM}~\cite{LoBeCe17} for loading BIM models from IFC files.

\section{Application Use Cases}
\label{sec:usecases}

In this section, we demonstrate application use cases where BEMtrace supports users in understanding BIM-to-BEM conversions. These use cases have been developed in close collaboration with the domain experts and cover typical analysis questions that arise when trying to understand the BIM-to-BEM conversion.

\begin{figure}[t!]
    \begin{tikzpicture}[outer sep=0, inner sep=0]
        \BeginAccSupp{method=pdfstringdef, ActualText={On the left side, a 3D BEM model is shown. Divided by a black bar, we see two more renderings showing the Relationship View. The first version shows a watertight model of the building. The second one shows the raw space boundary data where gaps between the planes can be seen.}}
            \node[anchor=north west] at (0,0) {\includegraphics[width=.98\columnwidth]{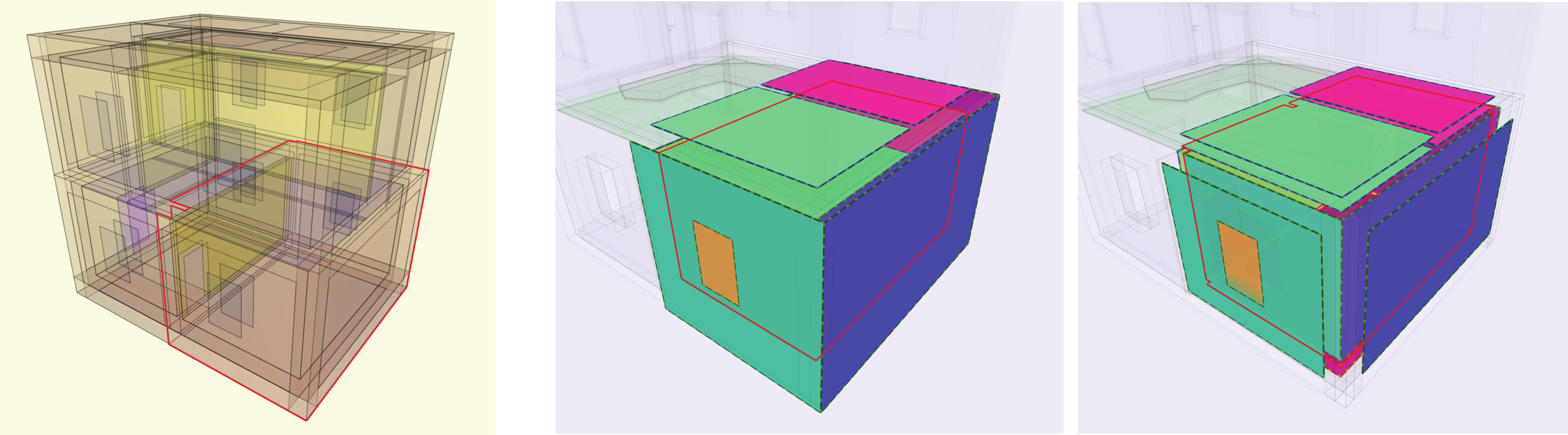}};
            \node[anchor=north west] at (0.6,-2.45) {\includegraphics[height=0.4cm]{pics/icon-bemview}};
            \draw (2.9,0) -- (2.9, -2.9);
            \node[anchor=north west] at (4.6,-2.45) {\includegraphics[height=0.4cm]{pics/icon-relationshipview}};
        \EndAccSupp{}
    \end{tikzpicture}
    \caption{Use Case 1: The users selected a room in the BEM View to check the related space boundaries in the Relationship View. Switching between the two options (\emph{Raw Data} and \emph{Enhanced}), users could see how the space boundaries had to be adapted (i.e., enlarged) by the conversion algorithm.}
    \label{fig:usecase1}
\end{figure}

\paragraph{Use Case 1: Successful BIM-to-BEM conversion} For the loaded BIM model, users started the BIM-to-BEM conversion without any issues or conflicts. However, the users were unsure whether a specific space boundary was correctly defined in BIM and then correctly transferred to a connection in BEM. To check this, the users selected a room in the BEM World from the tree view. Afterward, the users switched to the Relationship View to inspect the space boundaries. The users then iteratively switched between the raw data and the space boundaries that have been changed by the automatic BIM-to-BEM conversion. When comparing the raw space boundaries with the space boundaries after the conversion, users recognized that the algorithm needed to adjust the size of the boundaries to cover the whole room surface (illustrated in Figure~\ref{fig:usecase1}). By checking the Space View, users could confirm that the correct spaces had been connected. After inspecting the changes made by the BIM-to-BEM conversion, users accept the generated BEM model as a valid conversion.

\begin{figure}[b!]
    \begin{tikzpicture}[outer sep=0, inner sep=0, font=\sffamily\small]
        \BeginAccSupp{method=pdfstringdef, ActualText={This image consists of three 3D renderings showing the Space View. In the first view, all spaces are rendered as blue blocks. In the second one, the outdoor spaces are colored in green. In the last one, only the ground space representing the earth is shown, rendered as a brown block.}}
            \node[anchor=north west] at (0,0) {\includegraphics[width=.97\columnwidth]
            {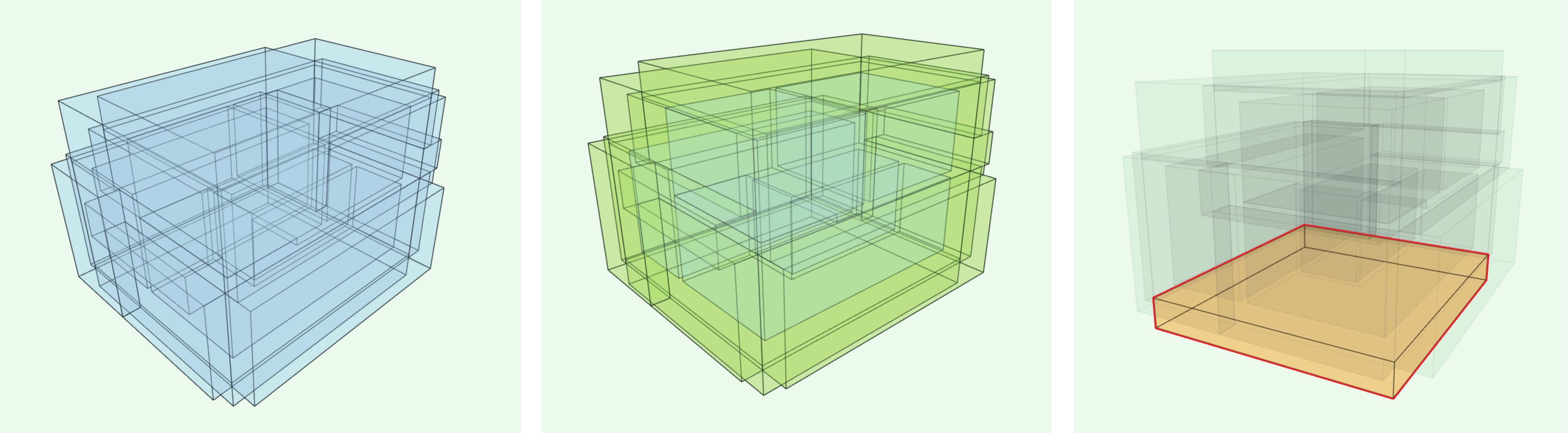}};
            \node[anchor=north west] at (.6,-2.45) {\includegraphics[height=0.4cm]{pics/icon-spaceview}};
            \node at (5.7,-2.7) {User corrections};
        \EndAccSupp{}
    \end{tikzpicture}
    \caption{Use Case 2: All spaces were incorrectly classified as internal ones. After user intervention, external spaces and the ground space could be correctly classified.}
    \label{fig:usecase2}
\end{figure}

\paragraph{Use Case 2: Resolving room classification issue} Loading another BIM model, the users started the BIM-to-BEM conversion, which was also completed without any problems. However, users needed clarification as to why external wall connections were not visible in the tree view of the generated BEM. To understand this issue, they opened the Space View and inspected the assigned room types. Here, they could see that the applied heuristics incorrectly identified all spaces as internal ones (Figure~\ref{fig:usecase2}). Consequently, users changed the heuristics so that external spaces could be identified correctly. In addition, users also selected the space representing the ground and classified it correspondingly. After recomputing the BEM model, all connections could be correctly computed.

\begin{figure}[t!]
    \begin{tikzpicture}[outer sep=0, inner sep=0]
        \BeginAccSupp{method=pdfstringdef, ActualText={The image on the left shows a 3D rendering of a building in the BEM View, with some windows marked in red and two in yellow. A 3D rendering of the same building in the Element View is shown on the right side. Arrows point outward from the two windows marked in yellow on the left side.}}
            \node[anchor=north west] at (0,0) {\includegraphics[width=.97\columnwidth]{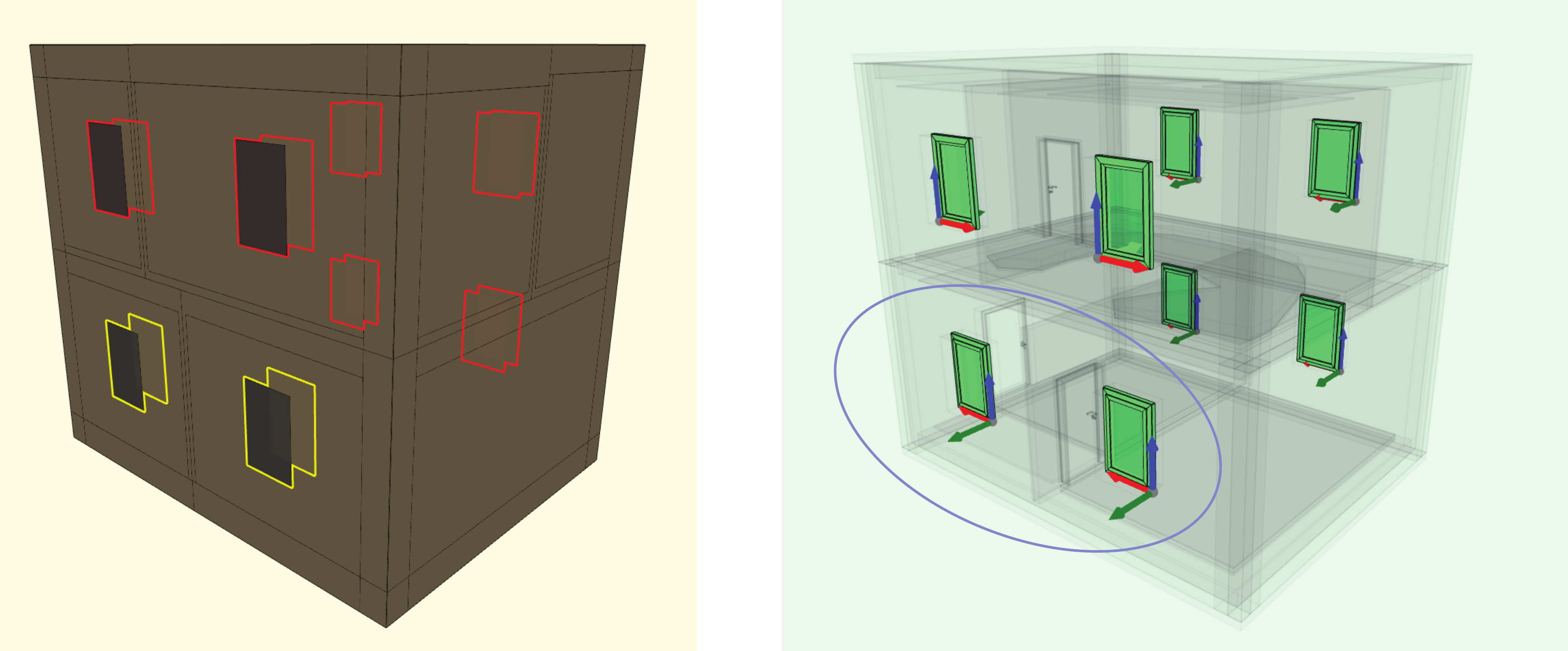}};
            \node[anchor=north west] at (1.2,-3.65) {\includegraphics[height=0.4cm]{pics/icon-bemview}};
            \draw (4.1,0) -- (4.1, -4);
            \node[anchor=north west] at (5.5,-3.65) {\includegraphics[height=0.4cm]{pics/icon-elementview}};
        \EndAccSupp{}
    \end{tikzpicture}
    \caption{Use Case 3: Two external windows have been incorrectly put into a second sub-group with different material properties. This happened due to the windows being oriented in a different way (green arrows pointing outwards), which can be seen in the Element View (blue circle).}
    \label{fig:usecase3}
\end{figure}

\paragraph{Use Case 3: Resolving issues in case of incorrectly grouped elements} In a BEM model, all connections to external rooms via windows were divided into two groups with different classifications. All windows leading to external rooms were classified as $AF$ (\emph{Außenfenster} in German, which can be translated as external windows), and within this group, the two sub-groups $AF1$ and $AF2$ have been generated. In Figure~\ref{fig:usecase3}, the two window sub-groups are marked in red and yellow. This confused the users, as all the windows in the building were supposed to have the same material properties. The users selected all external windows in the BEM View by utilizing the tree view and switched to the Element View. In the Element View, it is possible to show the local orientation of the elements. Here the users could see that two of the windows were placed in BIM with an incorrect orientation (blue circle in Figure~\ref{fig:usecase3}). The users corrected this and reloaded the BIM model.

\begin{figure}[b!]
    \begin{tikzpicture}[outer sep=0, inner sep=0]
        \BeginAccSupp{method=pdfstringdef, ActualText={The image shows a 3D rendering in the Relationship View. Two spaces, one indoor space, and one outdoor space are rendered as violet blocks. Another outdoor space below the other outdoor space is rendered as a red block. The indoor space is connected to both outdoor spaces.}}
            \node[anchor=north west] at (0,0) {\includegraphics[width=.97\columnwidth]{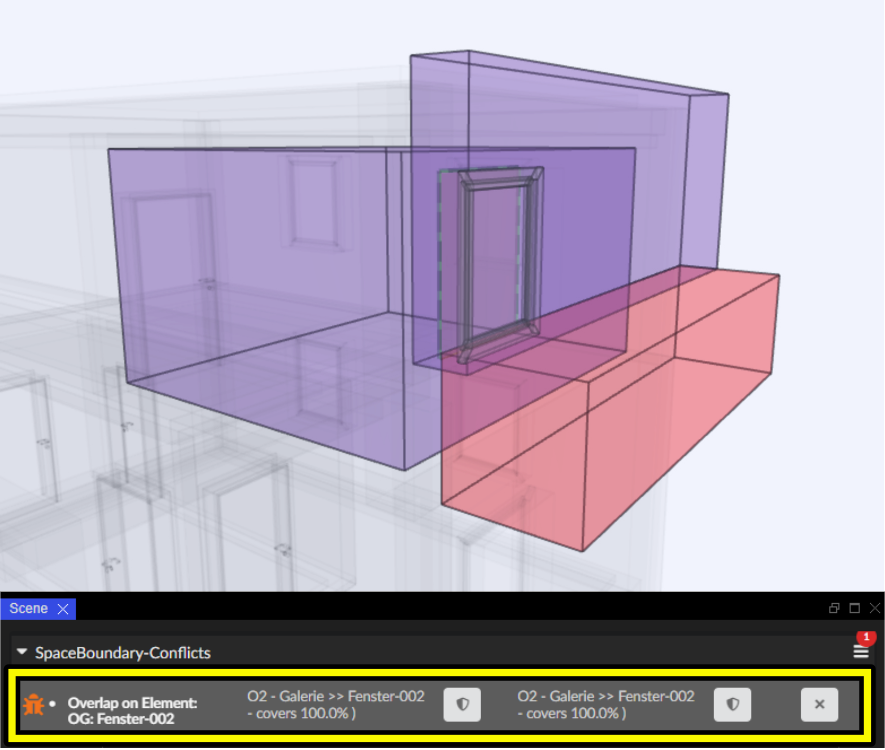}};
            \node[anchor=north west] at (3,-7.3) {\includegraphics[height=0.4cm]{pics/icon-relationshipview}};
        \EndAccSupp{}
    \end{tikzpicture}
    \caption{Use Case 4: In this conflict, a window is connected to two external spaces. Since thermal flow can only go from one space to another, the users had to manually decide which external space to keep.}
    \label{fig:usecase4}
\end{figure}

\paragraph{Use Case 4: Resolving a conflict} Geometry conflicts in the BIM data are automatically identified by the BIM-to-BEM conversion. Users are informed about conflicts in the Relationship View. Users can inspect and resolve conflicts directly in the interface or reload the BIM model. In this use case, an external window was connected to two external spaces. For inspecting this conflict, all required information was shown in the Relationship View. In Figure~\ref{fig:usecase4}, the users clicked on the conflict, and then the related spaces and element are shown in the Relationship View. Two pairs of overlapping space boundaries have been detected on the window surface, which would lead to an incorrect thermal flow. It was not possible to solve this conflict automatically since both resolutions were technically valid. Domain knowledge was needed to fix the issue. For example, one of the external spaces may have been created due to a planned balcony at this position - which is not essential for calculating energy transport in BEM. In this case, the users kept the upper external space and ignored the lower one in the BIM-to-BEM conversion.

\begin{figure}[t!]
    \begin{tikzpicture}[outer sep=0, inner sep=0]
        \BeginAccSupp{method=pdfstringdef, ActualText={The image on the left shows a 3D rendering in the Relationship View. In a two-story building, two spaces (one on the upper floor and one on the lower floor) are colored violet and orange blocks. A U-shaped floor element is rendered as grey 3D objects in the middle. The same building is shown as a 3D rendering on the right in the Element View. Here, the floor element is rendered as a green 3D object at the same position as the whole floor (violet block).}}
            \node[anchor=north west] at (0,0) {\includegraphics[width=.97\columnwidth]{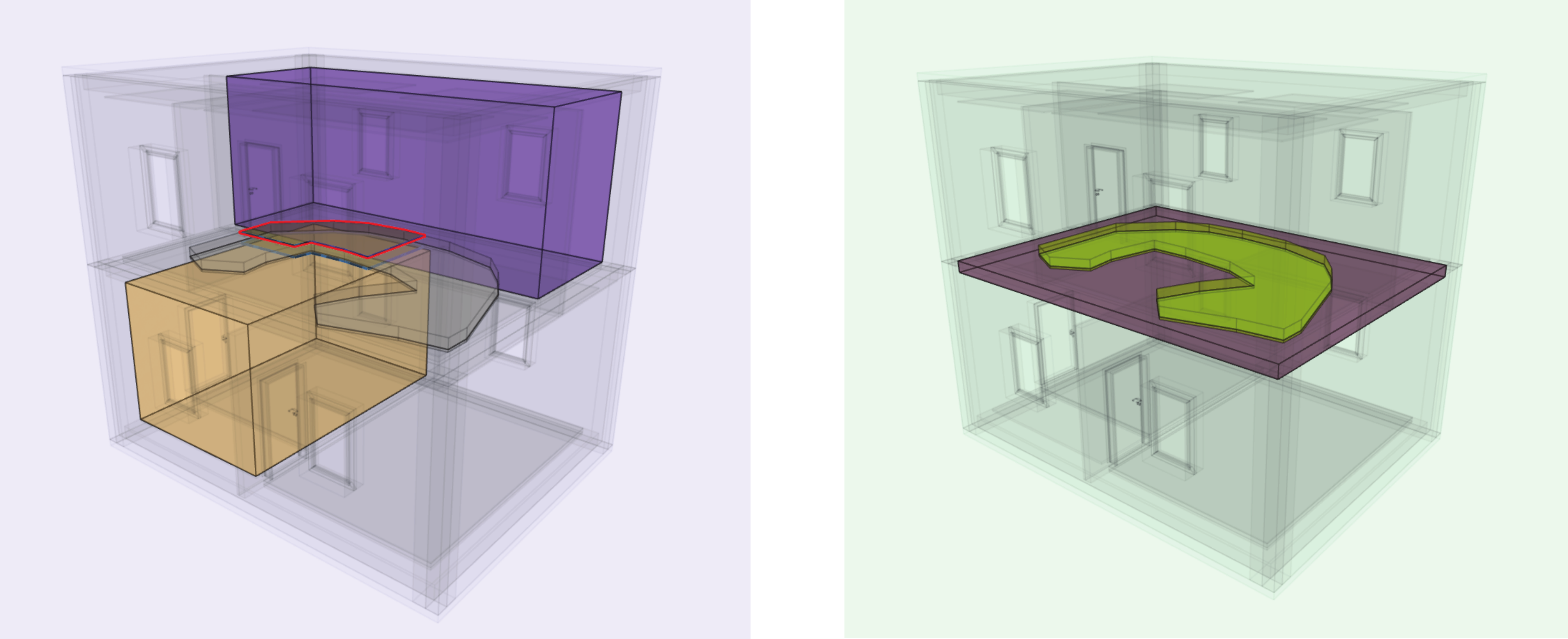}};
            \node[anchor=north west] at (.8,-3.55) {\includegraphics[height=0.4cm]{pics/icon-relationshipview}};
            \draw (4.45,0) -- (4.45, -3.9);
            \node[anchor=north west] at (5.5,-3.55) {\includegraphics[height=0.4cm]{pics/icon-elementview}};
        \EndAccSupp{}
    \end{tikzpicture}
    \caption{Use Case 5: In this conflict case, users had to solve a problem with overlapping space boundaries. The issue was caused by a slab that had been added to the BIM model just for testing purposes.}
    \label{fig:usecase5}
\end{figure}

\paragraph{Use Case 5: Resolving an artificial failure} In another BIM model, users were informed about a conflict due to several overlapping space boundaries. Inspecting the conflict in the Relationship View, users recognized that the problem was located between two floors. To better understand which elements are involved in the conflict, users switched to the Element View (illustrated in Figure~\ref{fig:usecase5}). Here they could see that one element has been drawn inside another. From their experience with the BIM model, the users knew that one of the elements, a slab with a rounded shape, was only added for testing purposes. Users could resolve the conflict by removing the slab from the BIM model.

\section{User Feedback}
\label{sec:evaluation}

To test BEMTrace regarding its abilities, strengths, weaknesses, and value, we combined four established evaluation methods to get broad feedback: (i) observational task performance analysis, (ii) thinking aloud~\cite{Lewi82}, (iii) heuristic-based evaluation (ICE-T)~\cite{WallAMDHES19}, and (iv) qualitative feedback. These methods are inexpensive to implement and do not require additional equipment (e.g., eye-trackers). According to the ICE-T method~\cite{WallAMDHES19}, only a small number of participants are required. Studies show that five evaluators are enough to catch more than $75\%$ of the problems. The applied evaluation process consisted of five consecutive parts:

\begin{enumerate}

    \item \textbf{Introduction:} We started the evaluation sessions by collecting necessary meta-data from the test person and introducing the user to the topic and evaluation procedure. We explained the general idea behind BIM, BEM, and BIM-to-BEM conversion and presented the different views and selection mechanisms. Participants had to agree to the evaluation’s privacy and anonymity terms.

    \item \textbf{Free exploration:} The test person was then able to explore the visualization freely and receive answers to any questions that arose. During this free exploration, participants were encouraged to ask questions and to think aloud. We took notes and collected user quotes.

    \item \textbf{Task analysis:} After getting familiar with the application, the participants had to solve two tasks. As a first task, we asked participants to select elements in BEM, trace their origins in the raw data in the other views, and explain how the raw data contributed or was changed to create the BEM model. As a second task, we asked participants to explain the conflict we described in Use Case 4 in Section~\ref{sec:usecases}.

    \item \textbf{Heuristic evaluation:} After completing the tasks, the test persons were asked to fill out the heuristic value-based survey (ICE-T) by Wall~et~al.~\cite{WallAMDHES19}.

    \item \textbf{Qualitative feedback:} We asked the participants to give feedback about the BEMTrace application and give suggestions for improvements.

\end{enumerate}

We created a simplified version of the BEMTrace application with fewer user interface elements to not distract participants. For example, we hid user interface elements to control some of the rendering parameters and domain-specific controls to change element parameters. The evaluation was conducted in person in a meeting room on a large display. Every evaluation took about $45$ to $60$ minutes. Two co-authors were present for every evaluation cycle, one explaining the application and the use cases, and the other one taking notes. We picked the following test users:

\begin{itemize}
    \item P1: age 41 years, female senior researcher and project leader, having a background in architecture and building physics.
    \item P2: age 28 years, male research engineer, having a Master's degree in computer science with a focus on computer graphics.
    \item P3: age 26 years, male research engineer, having a background in technical mathematics.
    \item P4: age 33 years, female researcher, having a Master's degree in computer science with a focus on computer graphics.
    \item P5: age 36 years, female researcher and project leader, having a Master's degree in computer science.
    \item P6: age 68 years, male civil engineer, having a background in building physics.
    \item P7: age 30 years, male engineer, having a background in energy technology.
\end{itemize}

\paragraph{Task analysis} P1, P3, P5, P6, and P7 quickly understood the context-adaptive selection mechanism and could soon switch between different views and explain what they saw. P2 and P4 needed some more advice on how to perform 3D selections. The Element View and Space View were quickly understood by all participants. All participants explored and understood the differences between the raw data and the algorithm's changes of the space boundaries in the Relationship View. P2, P4, and P5 needed time to differentiate between elements and the connected space boundaries. Most participants generally operated in the 3D view and only checked selections in the tree view. The option to render the building with a low-opacity silhouette was appreciated, as one could still see the parts of the building that had been filtered away. Labels have not been perceived as confusing. For analyzing the conflict case, we saw a difference between participants P1, P6, and P7, who had a background in working with 3D representations of buildings, and P2, P3, P4, and P5 with a computer science background. P1, P6, and P7 could easily and very quickly understand and resolve the conflict. P2, P3, P4, and P5 required more selection tools (tree view, hover) and views (Element View, Space View) before they could eventually explain that the window was connected to two spaces.

\paragraph{Heuristic evaluation} We used the questionnaire proposed by Wall~et~al.~\cite{WallAMDHES19} to quantitatively evaluate the BEMTrace application. In this evaluation scheme, participants rate $21$ statements on a $7$-point Likert scale ($1$ = strongly disagree, $7$ = strongly agree). Participants were allowed to skip questions if they felt they were not relevant to the application, which three participants did for individual questions. The average ratings of the questions, accumulated by topic, are given in Table~\ref{tab:evaluation}. An average value of greater than five is considered a success. We received good scores from all participants for Insight, Time, Essence, and Confidence. The lowest scores were given by P7, a domain expert who is accustomed to specialized visualizations in the field of energy technology. When he saw the visualizations for the first time, he felt that there was not enough time during the user study to become familiar enough with our visualizations to fully trust them.

\begin{table}[t!]
	\centering
    \caption{Evaluation results. The scores are according to the evaluation components \emph{Insight}, \emph{Time}, \emph{Essence}, and \emph{Confidence} as defined by Wall~et~al.~\cite{WallAMDHES19}. An average value greater than five is considered a success.}
    \label{tab:evaluation}
	\def\arraystretch{1.5}
    \BeginAccSupp{method=pdfstringdef, ActualText={The table consists of nine columns and five rows. The cells in the middle are colored according to the user study results. Results greater than 6.0 are green - the higher, the more saturated. Results between 5.0 and 5.9 are colored in orange. The one value for Confidence for P7 is colored in red.}}
    	\begin{tabular}{r|rrrrrrr|c}
    		& \textbf{P1} & \textbf{P2} & \textbf{P3} & \textbf{P4} & \textbf{P5} & \textbf{P6} & \textbf{P7} & \textbf{Avg.} \\ \hline
    		\textbf{Insight} & \cellcolor{EvaluationMiddle!64}6.4 & \cellcolor{EvaluationGood!69}6.9 & \cellcolor{EvaluationMiddle!60}6.0 & \cellcolor{EvaluationGood!70}7.0 & \cellcolor{EvaluationGood!70}7.0 & \cellcolor{EvaluationGood!70}7.0 & \cellcolor{EvaluationLow!53}5.3 & 6.5 \\
    		\textbf{Time} & \cellcolor{EvaluationMiddle!64}6.4 & \cellcolor{EvaluationGood!66}6.6 & \cellcolor{EvaluationGood!70}7.0 & \cellcolor{EvaluationMiddle!65}6.5 & \cellcolor{EvaluationGood!68}6.8 & \cellcolor{EvaluationGood!68}6.6 & \cellcolor{EvaluationLow!53}5.2 & 6.5 \\
    		\textbf{Essence} & \cellcolor{EvaluationLow!53}5.3 & \cellcolor{EvaluationGood!68}6.8 & \cellcolor{EvaluationGood!67}6.7 & \cellcolor{EvaluationMiddle!65}6.5 & \cellcolor{EvaluationGood!67}6.7 & \cellcolor{EvaluationMiddle!67}6.5 & \cellcolor{EvaluationMiddle!63}6.3 & 6.3 \\
    		\textbf{Confidence} & \cellcolor{EvaluationLow!58}5.8 & \cellcolor{EvaluationGood!70}7.0 & \cellcolor{EvaluationGood!70}7.0 & \cellcolor{EvaluationGood!68}6.8 & \cellcolor{EvaluationGood!70}7.0 & \cellcolor{EvaluationGood!68}6.8 & \cellcolor{EvaluationBad!40}4.0 & 6.3    
    	\end{tabular}
    \EndAccSupp{}
\end{table}

\paragraph{Qualitative feedback} During the final qualitative feedback round, participants stated that they generally understood the 3D representations and the selection mechanisms. Participants confirmed that they gained a better understanding of the origin of BEM parts and possible data issues when analyzing conflicts. Nevertheless, they also stated that '\emph{this is definitely a tool for experts, not for the general public}'. Some participants considered the filtering options for adjusting the rendering and the selection mechanisms as really necessary. P5 noted in this respect: '\emph{the Relationship View is a little bit confusing because there is a lot to see}'. We also received feedback that domain knowledge is required to correctly interpret the objects included in the BEM. P1 and P7 confirmed that '\emph{the division into different space boundaries is a novelty compared to existing tools}'. P6 emphasized trust building through visualization, because '\emph{the different views allow us to judge how the algorithm actually works. It enables me to trust this workflow and application}'.

%
%
%
%
%
%
%
%
%
%
%
%
%

\section{Discussion and Future Work}
\label{sec:discussion}

BEMTrace outlines a workflow for using visualization to understand a \textbf{3D data wrangling process}. The presented set of views connects views to either the raw data (BIM World), the conversion process (Relationship View), or the generated result (BEM World). Context-adaptive selections allow users to navigate through the different views. 
On a more general level, BEMTrace provides interlinked views and interactions on spatial data in successive states of processing. We argue that other domains dealing with 3D polygonal geometry and involving complex 3D data wrangling pipelines can find value in our approach. Such scenarios include point cloud alignment and reconstruction in mobile mapping and high-precision simulation and analysis tasks in digital twins.

The BEMTrace application still faces some \textbf{limitations}, which at the same time indicate directions for future work. 

Scalability in visualization~\cite{RicPAFSW24} can address different aspects. In BEMTrace, we identified \emph{algorithm scalability} (i.e., runtime performance) and \emph{parallel computing scalability} (i.e., parallelization capability) as minor issues. Although the BIM-to-BEM conversion algorithm scales only linearly with building size, users may wait for the computation to finish and continue the visual exploration of the result afterward. \emph{Cognitive and perceptual scalability} is also considered a minor challenge in BEMTrace since our domain experts are used to work with 3D representations and will become familiar with the visualizations if given more training time. \emph{Visual scalability} (i.e., visual clutter) can be identified as the greatest challenge for BEMTrace. To design and illustrate the workflow, we focused on small, two-story houses and rooms with rectangular layouts. Visualization complexity increases in the case of larger and more complex buildings. In practice, users prefer to partition work environments into small subsections of the building (e.g., floors), for which our visualizations are well suited. We provide filtering tools to facilitate the workflows, including geometric (i.e., floor-by-floor) and semantic (i.e., by some BIM attribute) selections.

In \textbf{future work}, we will explore options for visualizing the effects of our heuristics for selecting space types (see Use Case 2). We will improve the visibility of colors and labels in the 3D views. For example, in the Space View, rooms are colored according to type, but if many colored faces overlap, it is not easy to perceive an individual color in the view. Rendering parameters like transparency could be automatically adjusted according to the current selection.

The Relationship View, although considered overloaded by the computer scientists in our evaluation, attracted the most attention from the domain experts. In the future, we will especially concentrate to improve and further develop this view. BEMTrace will benefit from classification and a more structured visualization of the possible data issues and errors that might happen when transferring BIM to BEM. We would also like to improve the display of conflicts in the user interface (e.g., by grouping). As we noticed during the evaluation, users require additional guidance mechanisms (e.g., occlusion handling~\cite{WuPop16} or labels~\cite{EleJTL23}) to more easily become familiar with the visualizations. We continuously work with domain experts to further align BEMTrace with their needs and expectations.

The representation of the BEM model still needs improvement. In fact, BEM is based on a very dense spatial network specifying thermal flow between rooms. Depicting this network in a concise and easy-to-understand way will be a major challenge in the future. Visualization as a support tool for energy modeling has already been mentioned by Prouzeau~et~al.~\cite{ProuDBHHW18}. However, the spatial representation of the thermal flow is still an open research topic.

\section{Conclusion}
\label{sec:conclusion}

In this paper, we present BEMTrace, an application for understanding the complex 3D data wrangling process of converting a BIM (Building Information Modeling) model into a BEM (Building Energy Modeling) model. The paper showcases a visualization-driven approach for understanding 3D data conversion and introduces the domains of Building Energy Modeling and BIM-to-BEM conversions to the visualization community. We apply a novel technique with context-adaptive selections and visibility controls to support users in tracing back BEM parts to the original data stored in BIM. We evaluated the effectiveness of the BEMTrace application through various types of user feedback.

\acknowledgments{%
We thank the GEOSMAQ team, especially Christian Luksch and Martin Mautner, for their support and Herbert Raab for his trust in this project. We thank Stefan Bruckner for his valuable comments during the concept phase of the paper. VRVis is funded by BMK, BMAW, Styria, SFG, Tyrol and Vienna Business Agency in the scope of COMET - Competence Centers for Excellent Technologies (879730) which is managed by FFG.
}

\section*{Supplemental Material}

 The supplemental material includes more details about the proposed application:
\begin{itemize}
    \item \textbf{Demo video}: We illustrate the interactive operation on typical use cases in a demo video.
    \item \textbf{Evaluation details}: More information on the evaluation and the detailed scores of the ICE-T test are given in the supplemental material.
    \item \textbf{Full screenshot}: The entire application is shown on a full-screen capture.
\end{itemize}

\bibliographystyle{abbrv-doi-hyperref}

\bibliography{envis-vis-2024}

\end{document}